\documentclass[reprint,aps,prl,usenatbib,twocolumn,amssymb,superscriptaddress,nofootinbib,nobibnotes]{revtex4}

\usepackage{hyperref}

\usepackage{amsmath, amssymb}
\usepackage{esint}
\usepackage{graphicx}
\usepackage{subfigure}
\usepackage[english,english]{babel}

\usepackage{amssymb,amsfonts,textcomp}
\usepackage{supertabular}
\usepackage{hhline}
\usepackage{hyperref}
\usepackage[usenames]{color}
\usepackage{fancyvrb}
\usepackage{fancybox}

\definecolor{grey}{rgb}{0.75,0.75,0.75}
\definecolor{Orange}{rgb}{1.0,0.5,0.15}
\definecolor{brown}{rgb}{0.7,0.25,0.0}
\definecolor{pink}{rgb}{1.0,0.5,0.5}
\definecolor{darkerred}{rgb}{0.8,0,0}
\definecolor{darkerblue}{rgb}{0,0,0.8}
\definecolor{Blue}{rgb}{0,0.08,0.65}
\definecolor{Red}{rgb}{0.65,0.08,0.05}
\definecolor{Green}{rgb}{0.15,0.45,0.25}

\usepackage{color}
\usepackage{epsfig}
\usepackage{graphicx}
\DeclareGraphicsExtensions{.eps,.ps,.eps.gz,.ps.gz,.eps.Z,{}}

\usepackage{dcolumn}
\usepackage{bm}

\newcommand{\dd}{{\rm d}}

\newcommand{\mg}{\big<}
\newcommand{\md}{\big>}

\newcommand{\mC}{{\cal C}}

\newcommand{\mS}{{\cal S}}

\newcommand{\hw}{{\hat w}}

\newcommand{\hn}{{\hat n}}

\newcommand{\beq}{\begin{equation}}
\newcommand{\eeq}{\end{equation}}
\newcommand{\beqa}{\begin{eqnarray}}
\newcommand{\eeqa}{\end{eqnarray}}

\def\fun#1#2{\lower3.6pt\vbox{\baselineskip0pt\lineskip.9pt
        \ialign{$\mathsurround=0pt#1\hfill##\hfil$\crcr#2\crcr\sim\crcr}}}

\newcommand{\eff}{{\rm eff.}}

\newcommand{\simlt}{\lower.5ex\hbox{$\; \buildrel < \over \sim \;$}}

\newcommand{\mV}{{\cal V}}

\newcommand{\fsky}{f_{{\rm sky}}}
\newcommand{\nw}{{\rm nw}}
\newcommand{\bin}{{\rm bin}}
\newcommand{\matter}{{\rm matter}}

\thisfancyput(14.8cm,0.8cm){\large{YITP-20-15}}

\begin{document}
\title{Observing Baryonic Acoustic Oscillations in tomographic cosmic shear surveys}
\author{Francis Bernardeau} 
\affiliation{Institut d'Astrophysique
de Paris \& Sorbonne Universit\'e and CNRS, \\
98 bis boulevard Arago, 75014, Paris, France.}
\affiliation{Universit\'e Paris-Saclay, CNRS, CEA, Institut de physique th\'eorique, 
91191, Gif-sur-Yvette, France}
\author{Takahiro Nishimichi}
\author{Atsushi Taruya}
\affiliation{Center for Gravitational Physics, Yukawa Institute for Theoretical Physics, Kyoto University, Kyoto 606-8502, Japan}
\affiliation{Kavli Institute for the Physics and Mathematics of the Universe (WPI), UTIAS, The University of Tokyo, Kashiwa, Chiba 277-8583, Japan}
\date{\today}%
\begin{abstract}
We show that it is possible to build effective matter density power spectra in tomographic cosmic shear observations
that exhibit the Baryonic Acoustic Oscillations (BAO) features once a nulling transformation has been applied to the data.
The precision with which the amplitude and position of these features can be reconstructed is quantified in terms of sky coverage, 
intrinsic shape noise, median source redshift and number density of sources. BAO detection in Euclid or LSST like wide surveys 
will be possible with a modest signal-to-noise ratio.  It would improve dramatically for slightly deeper surveys.
\\
\end{abstract}
\maketitle

The shape deformation of background galaxies induced by the gravitational lensing of cosmological matter density fluctuations along the line-of-sight -- cosmic shear -- has been detected for the first time in the early 2000, \cite{2000A&A...358...30V,2000Natur.405..143W,2000MNRAS.318..625B}.
Results obtained  in more recent surveys (such as the CFHTLS survey \cite{2008A&A...479....9F}, DES\footnote{\url{https://www.darkenergysurvey.org}}, Subaru HSC \cite{2019PASJ...71...43H,2019arXiv190606041H}, KiDS (+VIKING) survey \cite{2018arXiv181206076H}) have confirmed that it is possible to build dedicated
cosmic shear wide surveys in which this effect can be measured with exquisite precision offering new means to constrain fundamental cosmological parameters.  This is at the heart of large projects such as the LSST\footnote{\url{http://www.lsst.org}} or EUCLID\footnote{see~\cite{2011arXiv1110.3193L}.}.

More precisely, on cosmological scale, cosmic shear signals depend on the geometry of Universe through a combination of angular
diameter distances, mass density of the universe and amplitude of the density fluctuations (see for instance \cite{1997A&A...322....1B}). With tomographic information, correlation functions in general give access to precise constraint on the expansion history of the Universe \cite{1999ApJ...522L..21H}. Another solid probe of the background expansion of the universe has been put forward in this context, namely the signature of sound waves in the photon-baryon plasma in the early Universe. It has now been measured with high precision in Cosmic Microwave Background (CMB) data\footnote{See for instance \cite{2014A&A...571A..16P,2016A&A...594A..14P} on how they have been used to constrain cosmological models and in particular dark energy parameters exploiting the fact that  the mechanisms at play are fully understood, e.g. \cite{1995PhDT..........H}.}. These Baryon Acoustic Oscillations (BAO) are however not limited to CMB observations: they have been seen in the galaxy distribution as a preferred co-moving separation of galaxies of ~150 Mpc \cite{2005ApJ...633..560E}, or, equivalently, as a series of oscillations in the galaxy power spectrum \cite{2005MNRAS.362..505C}. 

The purpose of the letter is to demonstrate that BAO features can also be detected in intrinsic cosmic shear observations, despite the fact that in essence weak lensing effects collect projected information of the large-scale mass distribution making it hard to detect the acoustic signature from the resultant featureless spectra. Here, we propose to use the nulling technique developed in \cite{2014MNRAS.445.1526B}, which can substantially mitigate the mixing of small- and large-scale modes due to projection effects, hence enables to make BAO features visible. We shall below examine in detail the feasibility of BAO detection based on this method.

In the context of this study we simply assume that cosmic shear observations lead to the construction of
multiple convergence maps $\kappa(\hat n)$ where $\hat n$ is a unit vector pointing in the celestial direction $\hat n$.
With detailed determination of the (photometric-)redshifts of the background galaxies, it is furthermore
possible to define vectors of cosmic-shear observations,
$\kappa_{i}(\hat n)$ corresponding to different populations of sources. In particular it is in principle possible to split the source populations in
redshift bins to create tomographic data sets as exemplified in \cite{1999ApJ...522L..21H}.

The data vectors we can manipulate are then convergence maps related to the mass density contrast, $\delta_\matter$, along the line of sight\footnote{This relation is however an approximation as it makes use of the Born approximation, ignores lens couplings and assumes the reduced shear can be  approximated by the shear itself. We do not think though that these approximations have significant impact on the developments presented in this paper.}:
\begin{equation}
\kappa_{i}(\hn)=\int_{0}\dd \chi\ w_{i}(\chi)\delta_{\matter}(\chi,\hn)
\end{equation}
where $w_{i}(\chi)$ is the resulting radial selection function for galaxies selected in the $i$-th redshift bin (written here as a function
of the comoving distance $\chi$), that is, in flat cosmology,
\begin{equation}
w_{i}(\chi)=\frac{3{\Omega_{0}H_0^2}}{2{c^2}}\int_{\bin_{i}}\dd z\ n(z)\ \frac{(\chi(z)-\chi)\chi}{\chi(z)a(\chi)}
\end{equation}
where $n(z)$ is the source redshift distribution function and the integration of $z$ is restricted to the bin $i$. The quantities $\Omega_0$, $H_0$ and $c$ are respectively the mass density parameter, Hubble parameter at present time, and the speed of light. The core observables are then the auto and cross-spectra of such maps given as a function of multipole $\ell$, $C_{ij}(\ell)$, defined as
\begin{equation}
\mg a^{(i)}_{\ell m} a^{(j)*}_{\ell'm'}\md=\delta_{\ell\ell'}\delta_{m m'}C_{ij}(\ell)
\end{equation}
where $a^{(i)}_{\ell m}$ are the spherical harmonics coefficients of the maps $\kappa_{i}(\hn)$. The cosmic shear spectra are then related to the (redshift dependent) matter power spectrum $P_{\rm m}(k,z)$ with
\begin{equation}
C_{ij}(\ell)=\int\frac{\chi'(z)\dd z}{\chi^{2}(z)}\,w_{i}(\chi(z))\,w_{j}(\chi(z))\,P_{\rm m}\left(\frac{\ell+1/2}{\chi(z)},z\right).
\end{equation}


\begin{figure}
   \centering
 \includegraphics[width=7cm]{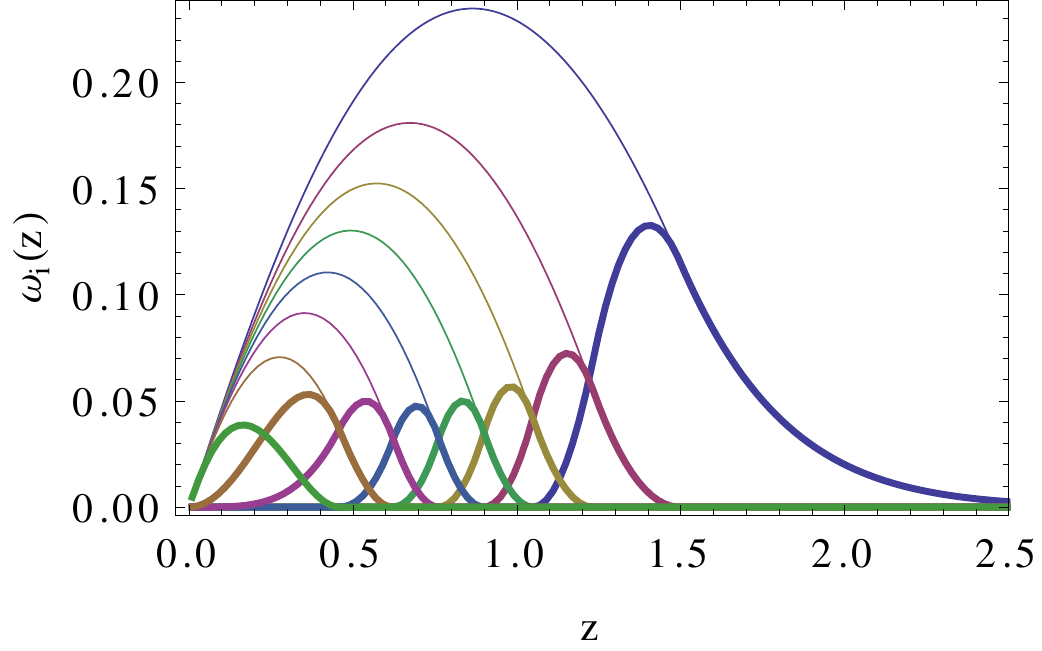}
   \caption{The nulling radial kernel functions. The thin lines are the standard tomographic weak-lensing kernel functions. The thick
   lines are those obtained after the nulling procedure has been implemented. The results are presented here for 8 equally populated
   tomographic bins.}
   \label{fig:winulling}
\end{figure}

The fundamental property we want to take advantage of is called nulling\footnote{Notions of nulling in this context were first introduced in \cite{2008A&A...488..829J}. We refer here to a more recent implementation of it.}.
It has indeed been shown in \cite{2014MNRAS.445.1526B} that one can define a transformation matrix $p_{ij}$ such that the transformed
radial selection function, $\hw_{i}=\sum_j p_{ij}w_{j}$, are fully localised in redshift. This is illustrated on Fig. \ref{fig:winulling}.
where the ``nulled''  radial selection function is shown as thick solid lines. We recall here how this transformation is built. Let us first define
the following quantities,
\begin{eqnarray}
n^{(0)}_{i}=\int_{\bin_i}\dd z\ n(z),\ \ n^{(1)}_{i}=\int_{\bin_i}\dd z\ n(z)\ \frac{1}{\chi(z)},
\end{eqnarray}
then, the two conditions (to be implemented for $i>2$),
\begin{eqnarray}
\sum_{j=i-2}^{i}{p_{ij}}\ n^{(0)}_{j}=0,\ \ \sum_{j=i-2}^{i}{p_{ij}}\ n^{(1)}_{j}=0
\end{eqnarray}
together with $p_{ij}=0$ for $j<i-2$ ensure that $\hw_{i}(z)$ vanishes for redshifts in the $1,\dots,i-3$ bins.

We illustrate this result in the case of a source distribution of the form,
\begin{equation}
n(z) \,\,{\propto}\,\, z^2 e^{-2^{3/4}\left(\frac{z}{z_{\rm m}}\right)^{3/2}}\label{ndez}
\end{equation}
with $z_{\rm m}$ is the median redshift of the sources which is about 0.9 for or the Euclid wide survey, \cite{2011arXiv1110.3193L}.
As stressed in \cite{2014MNRAS.445.1526B}, this transformation allows to separate the contributions of different scales. It was pointed in particular that it permits the application of the Perturbation Theory for the calculations of power spectra (see below). We show explicitly here that, by the same token, it allows to actually locate features in power spectra, such as BAO.

\begin{figure}
   \centering
 \includegraphics[width=6.5cm]{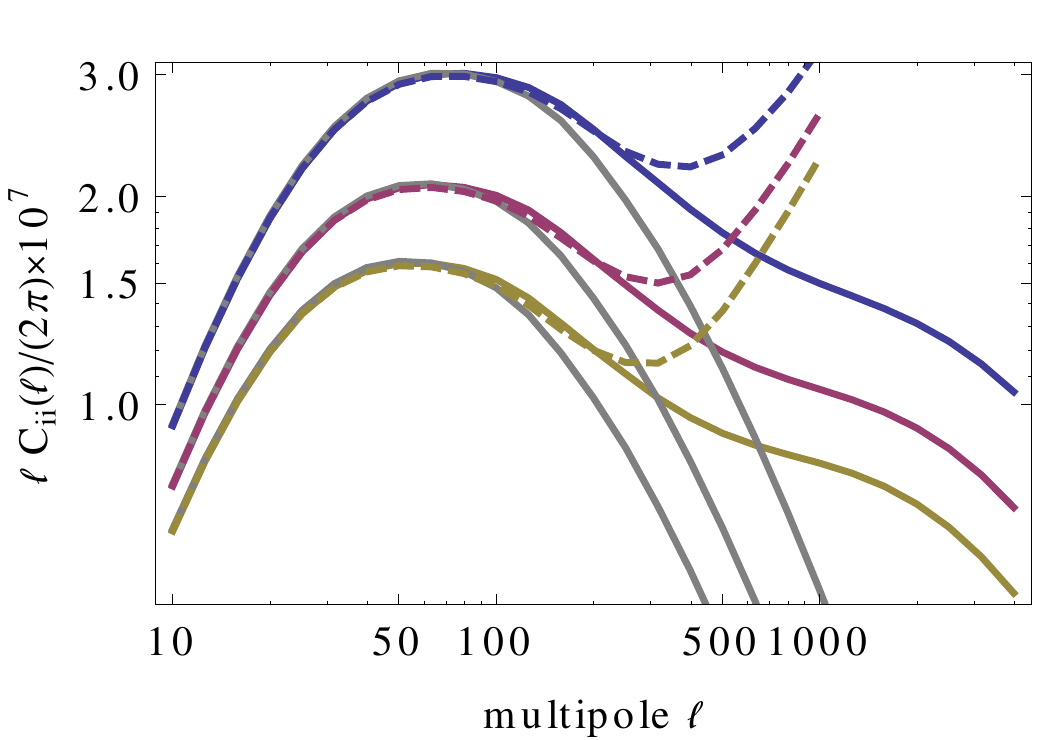}
 \includegraphics[width=6.5cm]{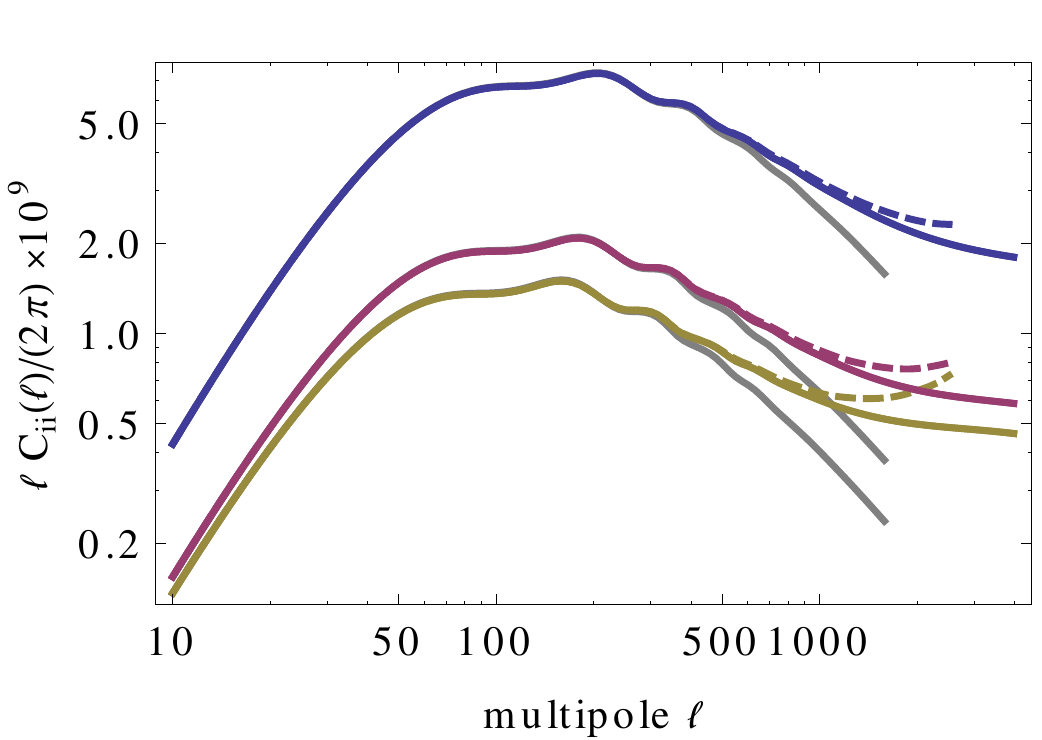}
   \caption{Cosmic shear tomographic (top panel) and nulled (bottom) spectra.
   The spectra are color-coded with the same convention as in Fig. \ref{fig:winulling}. The color lines correspond to the halofit power spectra. The solid and dashed grey lines correspond to linear and 2-loop standard perturbation theory prediction respectively. Most of the oscillatory features of interest here are well described by the  2-loop perturbation theory calculations which furthermore capture the departure, in amplitude and shape, from linear theory. We show here only the last three nulled bins for clarity.}
   \label{fig:lClBAO}
\end{figure}

More precisely the observables we want to exploit are defined as
\begin{equation}
\hat{C}_{ij}(\ell)=p_{ia}p_{jb}C_{ab}(\ell)\label{nulledspectra}
\end{equation}
and are built from the original spectra with the help of the nulling transformation coefficients.
The expected spectra $\hat{C}_{ii}$, for $i=6,7$ and $8$ are {shown}
on Fig. \ref{fig:lClBAO}. We show there the predictions for both the linear theory (gray solid lines), the next-to-next-to-leading (2-loop) order calculation (dashed lines)
of Standard Perturbation Theory (SPT) and halofit (colored solid lines). We refer here to the standard literature for the way SPT results are derived at such order \citep{2002PhR...367....1B,2013arXiv1311.2724B}. In practice it is computed via the RegPT code presented in \cite{2012PhRvD..86j3528T} which allows accurate and fast computation of the diagrams required at 2-loop order.
The halofit predictions  are based on the parameterization proposed in \cite{2003MNRAS.341.1311S} with the refined parameters by \cite{2012ApJ...761..152T}.
Clearly, with the nulling transformation, by restricting the range in comoving wavenumber $k$ which is spanned for a given $\ell$, the BAO features are made visible (bottom panel), and this is in marked contrast to what is obtained from standard tomographic spectra (top panel).

One can be more precise on the conditions for which such observations are made possible: for each $\ell$ and each bin $i$, the range in $k$ which is spanned for the computation of $C_{\ell}$ is now bounded. Its range is typically about
\begin{equation}
\Delta k_{\eff}(i;\ell)=\frac{\ell}{\chi_{\eff}(i)}\frac{\Delta\chi_{\eff}(i)}{\chi_{\eff}(i)},\label{Deltak}
\end{equation}
where $\chi_{\eff}(i)$ is the average angular distance of the bin $i$ and $\Delta\chi_{\eff}(i)$ its width.
One can check that
for $\ell$ about 200, $\Delta k_{\eff}(i)$ varies from $0.17$ to $0.005\,h/ $Mpc which, for the furthest bins, is indeed much smaller than the BAO wavelength in Fourier space (that is about $0.06\,h$/Mpc).
Furthermore, comparing the linear and the SPT or halofit  predictions shows that the BAO features are slightly smeared out because of non-linear couplings but  are  still clearly present. To be noted also is that although SPT has a limited validity range, it provides us with an accurate description of the BAO in the quasi-linear regime showing such features can then be captured from first principle calculations. Incidentally note that the range in which standard tomographic spectra can be accurately computed from first principle is much more limited.

One can take advantage of these quantities to construct effective power-spectrum estimators as a function
of the wavenumber $k$
\begin{eqnarray}
P_{\eff}^{(i)}(k)&=&\left({\int\frac{\dd\chi}{\chi^{2}}\,\frac{w_{i}^{2}(\chi)}{D_{+}^{2}(\chi)}}\right)^{-1}
{\hat{C}_{ii}(k\chi_{\eff}(i))}
\label{DefPeff}
\end{eqnarray}
where  $D_{+}(\chi)$ is the linear growth rate of the fluctuations. Such power spectra can then be aggregated together.

\begin{figure}
   \centering
 \includegraphics[width=7cm]{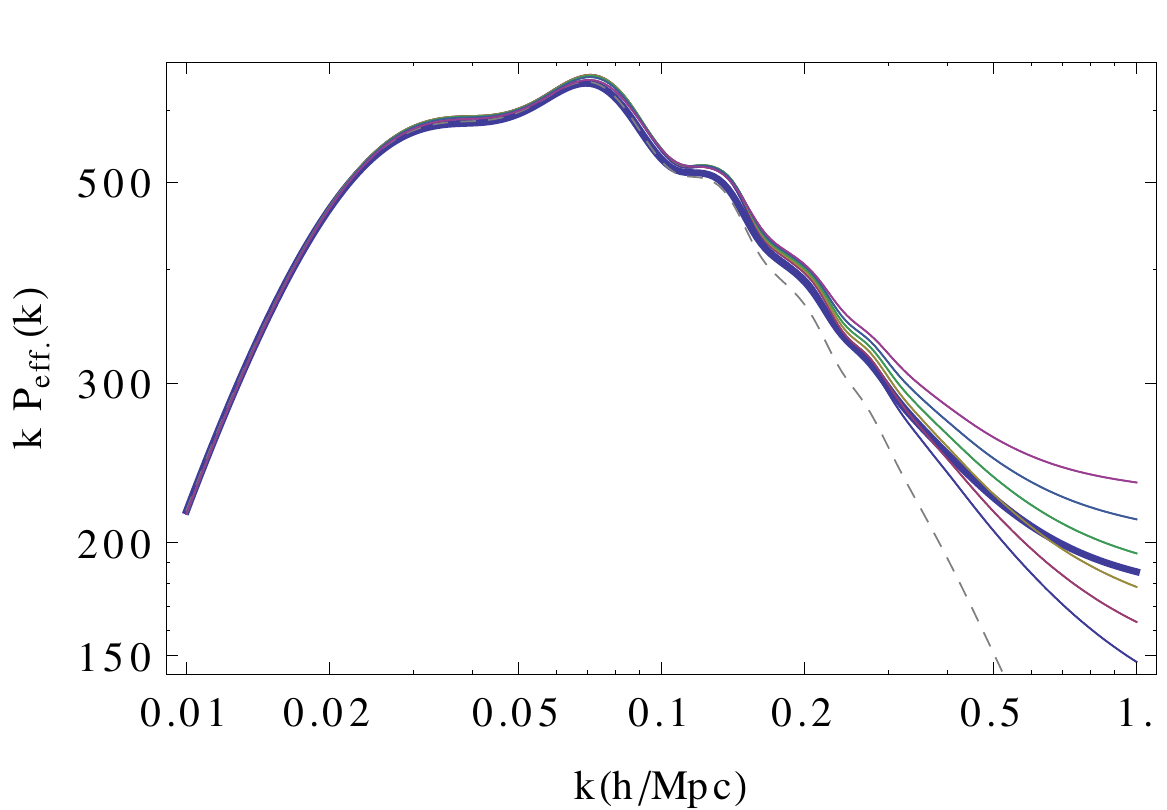}
   \caption{The reconstructed matter power spectra for each bin and from a combination of observed  cosmic shear spectra according to Eq. (\ref{DefPeff}).  The thin coloured lines are the power spectra from individual bins (where the nonlinear effects are all the more important that the bins are closer).  Bins 4 to 9 are shown assuming a 10 equi-populated binning. The grey dashed line is the prediction from the linear theory and the thick solid line is the reconstructed power-spectrum from the aggregation of bins 4 to 10.}
   \label{fig:RecPk}
\end{figure}

With such a construction it is then possible to coherently add  the observed spectra and preserve the information on the BAO positions.
In Fig. \ref{fig:RecPk} we present the result of such an exercise.
The adopted value for $z_{\rm m}$ is here 1.3. We used 10 bins to ensure the BAO features are sharp enough and use the bins 4 to 10 to do the reconstruction. Closer bins are found to slightly smear the oscillatory features. The end  result of the summation is an effective power spectrum whose shape is intermediate between the linear and nonlinear predictions as it aggregates observations obtained at different redshifts.

In the following we explore the precision with which one could constrain the amplitude and position of the BAO in realistic observational scenarios. Rather than $P_{\eff}(k)$ the starting point of the computation is to use the nulling power spectra and cross-spectra
as a data vector. The full data vector can then be defined as the collection of spectra
$\hat\mV_{ij}(\ell)$ for $i\in(1,n_{b})$, $j\ge i$, and we then assume to have
\begin{equation}
\hat\mV_{ij}(\ell)=\hat\mC_{ij}+\hat\mS_{ij}
\end{equation}
for the nulled channel neglecting other sources of systematics such as intrinsic alignment. The shape noise of the nulled spectra
is then
\begin{equation}
\hat\mS_{ij}(\ell)= p_{it}\ p_{jt}  \frac{\sigma_{s}^{2}}{n_{g}(t)},
\end{equation}
where $\sigma_{s}$ is the intrinsic shape r.m.s. -- and we adopt $\sigma_{s}=0.3$ \cite{2011arXiv1110.3193L} -- 
and $n_{g}(t)$ is the number density of galaxies in bin $t$. The data covariance is then computed assuming the 
field obeys Gaussian statistics so that all $\ell$ are independent and that each bin is obtained from the measurement 
of all available modes. More precisely, for a survey that covers a fraction $\fsky$ of the sky the number of available 
modes can be estimated to be about $(2\ell+1)\fsky$ as pointed in \cite{2007MNRAS.381.1018A}.
The covariance coefficients, $c_{ij;kl}(\ell)$, between data vector elements $\hat\mV_{ij}$ and $\hat\mV_{kl}$ are then
\begin{equation}
c_{ij;kl}(\ell)=\frac{1}{(2\ell+1)\fsky}\left[\hat\mV_{ik}(\ell)\hat\mV_{jl}(\ell)+\hat\mV_{il}(\ell)\hat\mV_{jk}(\ell)\right].
\end{equation}

The models we consider are simple single-parameter models for the linear density power spectra that either linearly interpolate between oscillatory and non-oscillatory models
or for two different positions of the oscillations. More specifically we define the $\alpha-$model,
\begin{equation}
\mC_{\alpha}(\ell)=\mC(\ell;\omega_{0}\!=\!-1)+\alpha\left[\mC_{\nw}(\ell;\omega_{0}\!=\!-1)-\mC(\ell;\omega_{0}\!=\!-1)\right]
\end{equation}
and the $\beta-$model,
\begin{eqnarray}
\hspace{-.4cm}
\mC_{\beta}(\ell)&=&\mC(\ell;\omega_{0}\!=\!-1)+\beta\frac{\dd}{\dd\omega_{0}}
\left[\mC(\ell;\omega_{0})-\mC_{\nw}(\ell;\omega_{0})\right]
\end{eqnarray}
where $\mC_{\nw}(\ell;\omega_{0})$ is the no-wiggle version of the spectra and where the projection effects are
computed for a dark energy component of constant equation of state $P=\omega_{0}\rho$.
The first model tests the possibility of detecting the oscillatory features at a given angular position ; the second tests
the possibility of determining their angular position through the dependence of the angular distance with the Dark Energy equation of state.
In our case, the posterior distribution of $\alpha$ (or $\beta$) behaves like,
\begin{equation}
p(\alpha)\sim \exp\left(-\frac{\alpha^{2} }{2}\ \sum_{\ell}I_{\alpha}(\ell)\right)
\end{equation}
where $I_{\alpha}(\ell)$ is a rate function for each $\ell$ obtained as
\begin{equation}
I_{\alpha}(\ell)=c^{-1}\cdot (\Delta\hat\mV) \cdot (\Delta \hat\mV)
\end{equation}
where $c^{-1}$ is the inverse matrix of $c_{ij;kl}$ and $\Delta\hat\mV$ is the difference between $\alpha=0$ and $=1$,
(or $\beta$).
The resulting r.m.s in the measurement of $\alpha$ (or $\beta$) is then
\begin{equation}
\sigma_{\alpha}=\left[\sum_{\ell=10}^{\ell_{\max}}I_{\alpha}(\ell)\right]^{-1/2}
\end{equation}
and we define the signal-to-noise ratio in the following as its inverse, {i.e., $(S/N)\equiv 1/\sigma_\alpha$ (or $1/\sigma_\beta$)}. 

\begin{figure}
   \centering
 \includegraphics[width=7cm]{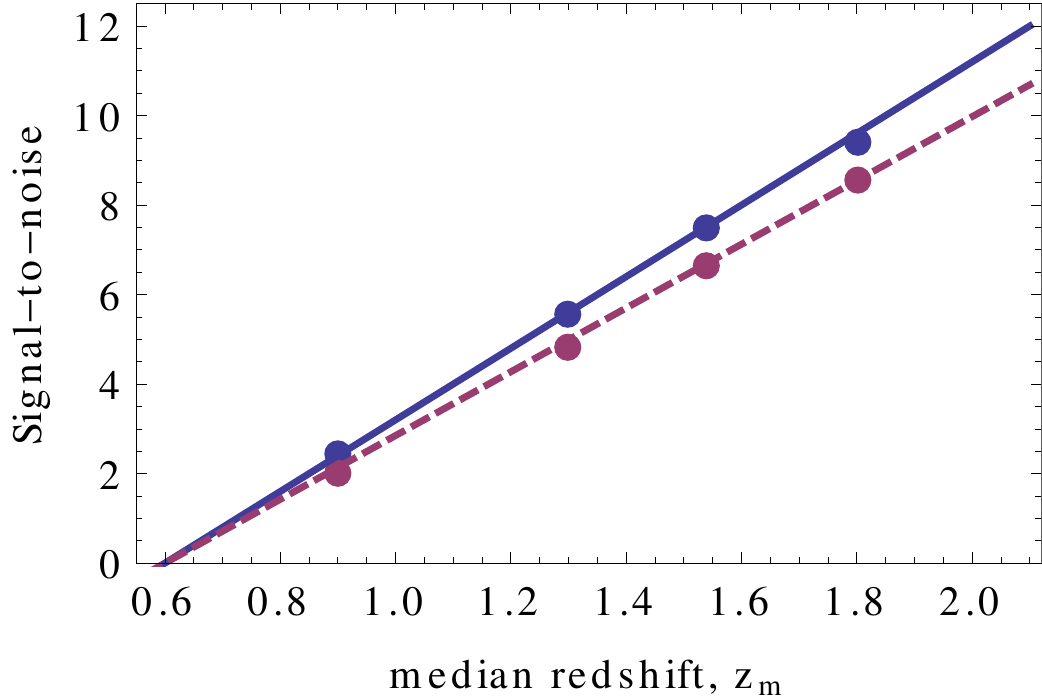}
   \caption{The dependence of the signal-to-noise ratio on observables $\alpha$ (solid blue lines) and $\beta$ (dashed red lines) on the median redshift when the number density of sources is changed accordingly. All but the first two bins are taken into account. The points are exact numerical results. The lines are fitted forms presented in Eq. (\ref{BAOSN}).}
   \label{fig:SignalToNoise}
\end{figure}

The signal-to-noise results are computed for a Euclid like mission with a sky coverage of 15.000 deg$^{2}$, assuming the source distribution of $z_{\rm m}=0.9$. The maximum multipole used for data analysis is set to $\ell_{\max}=850$ although the results depend very weakly on its choice and the adopted number of bins, $n_{b}=8$, corresponds to a case where the results have reasonably converged. No theoretical nuisance parameters are introduced as we are in a regime where power spectra can be computed from first principle.
The dependence of the signal-to-noise ratio with respect to the number density of galaxies or rather the depth
of the survey is then explored. It is obviously not straightforward to predict how the number density of galaxies -- and their $z$-distribution -- will evolve when one improves upon the sensitivity of the detection. For simplicity, while changing $z_{\rm m}$ we assume that the form (\ref{ndez}) generally holds and 
that the low $z$ amplitude of $n(z)$ is left unchanged naturally leading to a higher number density of galaxies for deeper surveys.
The results are presented in Fig. \ref{fig:SignalToNoise} showing that the signal-to-noise ratio regularly increases with $z_{\rm m}$ due to the combination of two effects: as the number of tracers increases the shape noise diminishes making the number of useful modes larger and the depth of the survey increases making the signal larger. These two effects are of similar amplitude. The resulting scaling of the signal-to-noise ratio is encapsulated in the form,
\begin{equation}
\left(\frac{S}{N}\right)_{{\rm BAO}}=3.2\left(\frac{\fsky}{0.375}\right)^{0.5}\left(\frac{\sigma_{\epsilon}}{0.3}\right)^{-1}
\left(\frac{z_{\rm m}-0.6}{0.4}\right)
\label{BAOSN}
\end{equation}
for the $\alpha$-model (and 3.2 is replaced by 2.85 for the $\beta$-model).
It shows that a significant signal-to-noise ratio can be obtained if the number density of galaxies is large enough. In practice
a non-ambiguous detection of the BAO would be secured if one pushes the median redshift to $z_{\rm m}=1.3$ using up to 90 gal/arcmin$^{2}$ in a wide survey. The performances of the  Euclid mission as it is designed \cite{2011arXiv1110.3193L}, with about 30 gal/arcmin$^{2}$  for a median redshift of 0.9,
offer then only a marginal detection. Ground based observations such as provided by the LSST project might perform slightly better with a median redshift of 1.2 but with only 30 gal/arcmin$^{2}$ available for cosmic-shear measurement, \cite{2019ApJ...873..111I}, relying on an efficient de-blending strategy \cite{2013MNRAS.434.2121C}. But relation (\ref{BAOSN}) clearly shows that observing BAO in cosmic shear is within reach of dedicated wide cosmic shear surveys.

\section*{Acknowledgments}
This work is supported in part by MEXT/JSPS KAKENHI Grants, Numbers JP15H05889 and JP16H03977 (AT). T. N. was supported by Japan Science and Technology Agency CREST JPMHCR1414, and by JSPS KAKENHI Grant Number JP17K14273 and JP19H00677.

\bibliographystyle{apsrev}
\bibliography{Nulling}
\end{document}